\documentclass[aps,twocolumn,prd,showpacs,nofootinbib]{revtex4}
\usepackage{amsmath}
\usepackage{graphicx}
\usepackage{dcolumn}
\usepackage{bm}
\usepackage{amssymb}
\usepackage{latexsym}

\begin{document}

\title {Supersymmetric oscillator: novel symmetries}

\author{R. Kumar$^{(a)}$}
\email{raviphynuc@gmail.com}
\author{R. P. Malik$^{(a, b)}$}
\email{malik@bhu.ac.in}
\affiliation {{\it $^{(a)}$Physics Department, Centre of Advanced Studies,}
{\it Banaras Hindu University, Varanasi - 221 005, India}\\
{\it $^{(b)}$DST Centre for Interdisciplinary Mathematical Sciences,}
{\it Faculty of Science, Banaras Hindu University, Varanasi - 221 005, India}}

\vskip 5cm

\begin{abstract}
We discuss various continuous and discrete symmetries of the supersymmetric simple
harmonic oscillator (SHO) in one (0 + 1)-dimension of spacetime and show their relevance
in the context of mathematics of differential geometry. We show the existence of a novel
set of discrete symmetries in the theory which has, hitherto, not been discussed in the
literature on theoretical aspects of SHO. We also point out the physical relevance of
our present investigation.

\end{abstract}



\maketitle

\section{Introduction}
The model of a harmonic oscillator (HO) is one of the most studied models in the history of
theoretical physics. This is mainly due to the fact that it is an exactly solvable cute model which
encompasses in its folds a rich and elegant mathematical structure. Theoretically, this
model has been able to explain a multitude of phenomena in diverse domains of physics
as well as other key branches of science. A supersymmetric version of the above HO,
that incorporates the bosonic and fermionic variables, provides a prototype example
of supersymmetry  and its innate supersymmetric algebra \cite{w,cas}.

The purpose of our present investigation is to discuss the discrete and continuous symmetries
of the one (0 + 1)-dimensional (1D) supersymmetric harmonic oscillator (SHO) and establish that
it is a cute 1D model for the Hodge theory. In fact, the symmetries of the SHO provide a physical
realization of the de Rham cohomological operators of differential geometry \cite{smm,egh}. Whereas the
continuous symmetries (and corresponding generators) provide the analogue of the de Rham cohomological
operators, the discrete symmetry of this model corresponds to the Hodge duality
operation of differential geometry. Thus, the SHO is a complete model for the Hodge theory.

In our earlier set of works \cite{r,sr1,sr2}, we have shown that the 1-form and 2-form gauge field theories
(in two (1+1)-dimensions and four (3 + 1)-dimensions of spacetime) provide 
physical models for the Hodge theory.
We have also discussed a toy model of a 1D rigid rotor and demonstrated that it provides a cute
model for the Hodge theory (at the algebraic level) \cite{sr2}. In all the above models, it is the existence of 
continuous and discrete symmetry transformations (and their corresponding generators) that provide the physical realizations 
of the abstract mathematical objects of differential geometry.

It is worthwhile to mention  that the above kind of studies have enabled us to prove that the 2D
Abelian 1-form and 4D Abelian 2-form gauge theories provide a new model for the topological field 
theory \cite{rpm1} and an example of the quasi-topological field theory \cite{rpm2}, respectively. 
Thus, such kind of studies are physically interesting.

All the above models are, however, based
on the gauge symmetries that are generated by the first-class constraints in the language of Dirac's
prescription for the classification scheme of constraints \cite{d}. So far, we have {\it not} studied a supersymmetric
model in the purview of mathematical structure of a Hodge theory. In our present investigation,
we try to accomplish this goal by taking the example of SHO where the physical continuous and discrete symmetries of
the theory play very important roles. We plan to extend our present work to other
superpotentials (e.g. shape-invariant potentials) which
have been shown to be of some physical interests \cite{cas}.

The following factors have contributed to our curiosity in pursuing our present investigation. 
First and foremost, the model of SHO is one of the prototype examples of a supersymmetric theory
which  has been studied from many different angles. Thus, it is always challenging to state something
{\it new} about this model. Second, to prove a model to be an example of the Hodge theory, one
has to examine and explore various kinds of symmetries so that the abstract mathematical quantities,
associated with the Hodge theory, could be explained in terms of the symmetry transformations. 
This is {\it an uphill} task. Thus, we are highly motivated to accomplish this goal with the
sophistication of theoretical physics. Finally, the present model is {\it not}
a gauge field theoretic model. Thus, the model of SHO is
unique in its own right because, even though, it is a model for the Hodge theory, it does not  
lean heavily on the idea of Dirac's first-class constraints and associated local and continuous gauge symmetries.

The contents of our present paper are organized as follows. To set up the notations and conventions, we 
recapitulate the bare essentials of the Lagrangian and Hamiltonian formulations of the SHO in section two. 
Our section three deals with the existence of a couple of 
nilpotent continuous symmetry transformations plus a bosonic 
symmetry transformation. The latter is obtained from the above nilpotent symmetry transformations. Our 
section four is devoted to the description of  discrete symmetries in the theory. We deduce the algebraic structures 
of the symmetry transformations (and corresponding generators) in section five. Finally, we 
make some concluding remarks in section six.

\section{Preliminaries: canonical formalism}  
We begin with the Lagrangian for the 1D SHO with unit mass (i.e. $m = 1$) and
natural frequency $\omega$. This  physical
system is described by the ordinary and Grassmannian 
dynamical variables at the classical level. The explicit form of the Lagrangian is (see, e.g. \cite{das})
\begin{eqnarray}
L = \frac {1}{2} \; {\dot x}^2 - \frac {1}{2} \; \omega^2 x^2 + i \; \bar \psi \;\dot \psi - \omega\; \bar \psi \;\psi, 
\end{eqnarray}
where $\dot x = ({dx}/{dt})$ and $\dot \psi = ({d\psi}/{dt}) $  are the generalized velocities of the SHO in terms of
its instantaneous position $x$, Grassmannian variable $\psi$ and the evolution parameter $t$. Here the pair
($\psi, \bar \psi$) are the Grassmannian variables (with $\psi^2 = 0, \bar \psi^2 = 0,  \psi \bar \psi + \bar \psi \psi = 0$) and 
we adopt the convention of  left derivative for these objects. As a consequence, we obtain the following Hamiltonian 
by exploiting the Legendre transformation, namely;   
\begin{eqnarray}
H &=& \dot \psi \;\;\Pi_{\psi} + \dot {\bar \psi}\; \;\Pi_{\bar \psi} + \dot x\; p - L\nonumber\\
&=& \frac {1}{2} \; p^2 + \frac {1}{2} \; \omega^2 \;x^2 + \omega \; \bar \psi\; \psi, 
\end{eqnarray}
where $p = ({\partial L}/{\partial \dot x}) = \dot x, \Pi_{\psi} = ({\partial L}/{\partial \dot \psi}) = - i \bar \psi, 
\Pi_{\bar \psi} = ({\partial L}/{\partial \dot {\bar \psi}}) = 0$
are the canonical conjugate momenta corresponding to the variables $x, \psi,$ and $\bar \psi$.

We can define the following bosonic and fermionic creation and annihilation operators in terms of the 
suitable dynamical variables of the Lagrangian. These operators (with $\hbar = c = 1$ and $m = 1$) are
(see, e.g. \cite{das})
\begin{eqnarray}
&&a_B^\dagger = \frac {1}{\sqrt{2\omega}}\;(- i \; p + \omega \;x), \qquad a_F^\dagger = \bar \psi,\nonumber\\
&&a_B = \frac {1}{\sqrt{2\omega}}\;(i\; p + \omega \;x), \qquad \;\;\;a_F = \psi.
\end{eqnarray}
In view of the above, we can easily check that
\begin{eqnarray}
H = \omega \;(a_B^\dagger \; a_B + a_F^\dagger \;a_F)
\equiv  \frac {1}{2} \; p^2 + \frac {1}{2} \; \omega^2 x^2 + \omega\; \bar \psi \psi.
\end{eqnarray}
We can verify, in a straightforward manner, that the following operators
\begin{eqnarray}
N_B = a_B^\dagger a_B, \;N_F = a_F^\dagger a_F, \; Q = a_B^\dagger a_F, 
\; \bar Q = a_F^\dagger  a_B,
\end{eqnarray} 
are the conserved quantities because they commute with the Hamiltonian of the theory if we use the 
following basic commutator and anticommutator brackets
\begin{eqnarray}
[a_B, a_B^\dagger ] = 1, \qquad \qquad \{a_F, a_F^\dagger \} = 1,
\end{eqnarray}
and take all the rest of the brackets to be zero. In other words, we take $({a_F^\dagger})^2 = \frac{1}{2}\; \{ a_F^\dagger, a_F^\dagger \} = 0,
a_F^2 = \frac {1}{2}\;\{ a_F, a_F \} = 0, [ a_B, a_F] = 0, [a_F^\dagger, a_B^\dagger ] = 0$, etc. 
Exploiting these brackets,
it can be proved that the conserved fermionic and bosonic 
quantities $Q, \bar Q, N_B, N_F$ and $H$ obey the following explicit algebra 
\begin{eqnarray}
&&\bigl[Q,\; H\bigr]\; = \;[\bar Q, \; H] = 0, \quad [N_B, \; H] \;=\; [N_F, \; H] = 0, \nonumber\\
&&Q^2 = \displaystyle{\frac {1}{2}\; \{Q, \; Q\} = 0}, \;\qquad {\bar Q}^2 = \frac {1}{2}\; \{\bar Q, \; \bar Q\} = 0,\nonumber\\ 
&& \{Q, \; \bar Q\} = \frac {H}{\omega},\quad
[Q, \;N_B] =  - Q, \quad [\bar Q, \;N_B] = \bar Q,\nonumber\\
&&[Q, \;N_F] =  Q, \quad [\bar Q, \;N_F] = - \bar Q,
\end{eqnarray}
which shows that the Hamiltonian $H$ of the theory  is the Casimir operator for the whole algebra.

We wrap up this section with the remarks that the following fermionic ($Q^2 = {\bar Q}^2$ = 0) conserved quantities
(i.e. $\dot Q = - i [Q, H] = 0, \dot {\bar Q} = - i [\bar Q,  H] = 0$), expressed in terms of 
the dynamical variables, namely;
\begin{eqnarray}
Q =  \frac {1}{\sqrt{2\omega}}\;(- i p + \omega x) \; \psi, \quad
\bar Q = \frac {1}{\sqrt{2\omega}}\; \bar \psi \;(i p + \omega x),  
\end{eqnarray}
can be derived from the nilpotent continuous symmetries of the Lagrangian (1) as the Noether conserved 
charges. Similarly, the Hamiltonian $H$ [cf. (4)] can also be derived as a conserved charge corresponding 
to a continuous bosonic symmetry (that is obtained from the above nilpotent symmetries). We discuss these 
continuous symmetry transformations in our next section.

\section{Continuous symmetries} 
It is interesting to note that under the following infinitesimal, local, continuous and 
nilpotent (i.e. $s_1^2 = 0, s_2^2 = 0$) 
transformations 
\begin{eqnarray}
s_1 x = \frac {- i\;  \psi}{\sqrt {2\omega}},\quad s_1 \bar \psi 
= \frac {1}{\sqrt {2\omega}} \; (\dot x + i\;\omega \;x ), \quad s_1 \psi = 0,\nonumber\\ 
s_2 x = \frac {i\;  \bar \psi}{\sqrt {2\omega}},\quad s_2 \psi 
= \frac {1}{\sqrt {2\omega}} \; (- \dot x + i\;\omega\; x ), \quad s_2 \bar \psi = 0,
\end{eqnarray}
the Lagrangian of SHO transforms as:
\begin{eqnarray}
s_1 L =  \frac {d} {dt}\Big(- \;\frac {\omega}{\sqrt{2\omega}}\; x \; \psi\Big),  \quad
s_2 L =  \frac {d} {dt}\Big(\frac {i}{\sqrt{2\omega}}\; \dot x \; \bar \psi\Big). 
\end{eqnarray}
As a consequence, the action remains invariant under the continuous transformations $s_1$ and $s_2$.
Now, there are two side remarks in order. First, the symmetry transformations
$s_1$ and $s_2$ are nilpotent of order two (i.e. $s_1^2 = 0, s_2^2 = 0$)
only on the {\it on-shell} where the equations of motion 
$\dot \psi + i\; \omega \psi = 0$ and $\dot {\bar \psi} - i\; \omega \bar \psi = 0$ are valid. Second,
the fermionic transformations $s_1$ and $s_2$ commute with the bosonic pair $(x, p)$ and anticommute with
the fermionic pair $(\psi, \bar\psi)$. The above inputs are important for our rest of the discussions in
our present endeavor.

Using the Noether's theorem, it is straightforward to verify that the conserved charges (8) emerge from the nilpotent
continuous symmetry transformations (9). In other words, 
the charges (8) are the generators of the symmetry transformations (9)
[for the Lagrangian (1) of SHO] because
\begin{eqnarray}
s_1 \Phi = \pm i\; [\Phi, \; Q]_{\pm}, \quad \qquad s_2 \Phi = \pm i\; [\Phi, \; \bar Q]_{\pm},
\end{eqnarray}
where $(+)-$ signs on the square bracket correspond to 
the (anti)commutator for the generic dynamical variable 
$\Phi = x, \psi, \bar \psi$ being (fermionic) bosonic in nature. Furthermore, 
depending on the (fermionic) bosonic nature of $\Phi$, we choose
the $(+)-$ signs, that are present,  in front of the square brackets in equation (11).

The nilpotent ($s_1^2 = 0, s_2^2 = 0$) transformations $s_1$ and $s_2$ do not anticommute (i.e. $\{s_1, \; s_2 \} \ne 0$). 
As a consequence, we define a bosonic symmetry transformation $s_W = \{s_1, \; s_2\}$ (i.e. $s_W^2 \ne 0$), under which, the dynamical 
variables ($x, \psi, \bar \psi$) transform as given below:
\begin{eqnarray}
&& s_W \;x = \frac {i}{\omega}\; \dot x, \qquad s_W \;\psi = \frac {i}{2\;\omega}\;(\dot \psi - i \; \omega \; \psi), \nonumber\\
&& s_W \;\bar \psi = \frac {i}{2 \;\omega}\;(\dot{\bar \psi} + i \; \omega \; \bar \psi).
\end{eqnarray}
The above transformations are {\it symmetry} transformations because the Lagrangian $L$ [cf. (1)] transforms 
to a total time derivative:
\begin{eqnarray}
s_W L = \frac {d}{dt}\; \Big[\frac {i}{2\;\omega}\;\Big({\dot x}^2 - \omega^2 x^2 
+ i \;\bar \psi\; \dot \psi -\omega \;\bar \psi\; \psi\Big)\Big].
\end{eqnarray} 
As a consequence, the action integral $S = \int dt \; L$ remains invariant for 
the physically well-defined dynamical variables that are present in the theory.

According to  Noether's theorem, we have the following expression for the conserved charge ($W$) corresponding to
the transformations $s_W$ in (12):
\begin{eqnarray}
W &=& \frac {i}{\omega} \; \Big(\frac {1}{2} \; {\dot x}^2 
+ \frac {1}{2} \; \omega^2 x^2 + \omega\; \bar \psi \;\psi\Big)\nonumber\\
&\equiv & \frac {i}{\omega} \; \Big(\frac {p^2}{2} 
+ \frac {1}{2} \; \omega^2 x^2 + \omega\; \bar \psi \;\psi\Big).
\end{eqnarray}
Thus, we note that the above conserved charge [i.e. $W = (i/\omega) H$] is connected with the Hamiltonian $H$
of SHO. As a consequence, basically 
$H$ is the generator of the infinitesimal, local and continuous transformations (12).
In reality, it is elementary to verify that $s_W \Phi = - i [\Phi, W]$ for $\Phi = x, \psi, \bar \psi$
when one uses the equations of motion $\dot \psi + i \omega \psi = 0, \dot {\bar \psi} - i \omega \bar\psi = 0$
derived from (1).

We close this section with the remark that the conserved operators $N_B = a_B^\dagger a_B$ and 
$N_F = a_F^\dagger a_F$, expressed in terms of the dynamical variables, also generate continuous symmetry
transformations for the Lagrangian (1). However, the ensuing symmetries are incorporated in the
symmetry generated by $W = (i/\omega) H \equiv (i/\omega)( a_B^\dagger a_B + a_F^\dagger a_F)$. 
This is precisely the reason
that these symmetries have not been discussed separately and independently in our present discussion.

\section {Discrete symmetries}
The Lagrangian $L$ of SHO also respects the following discrete symmetry transformations
%(besides the continuous symmetry transformations of Sec. 3).
%These transformations are illustrated below in explicit form, namely;
\begin{eqnarray}
\psi \to \pm \;i\; \bar \psi, \quad \bar \psi \to \mp\; i\; \psi, \quad x \to - x,\quad \omega \to - \omega.
\end{eqnarray} 
It is straightforward to check that the Lagrangian ($L$) remains invariant (i.e. $L \to L$) 
under the above discrete transformations. The transformations (15) are the analogue of the Hodge 
duality operation of differential geometry. To corroborate the above 
statement, first of all, it can be verified that two successive transformations, corresponding to (15), 
on the dynamical variables of the theory (i.e. $x, \psi, \bar\psi$) lead to the following \cite{dght}:   
\begin{eqnarray}
*\;(*\;\psi) = +\; \psi, \quad  *\;(*\;\bar \psi) =  +\; \bar  \psi, \quad  *\;(*\; x) = + \; x,
\end{eqnarray} 
where $(*)$ is the discrete symmetry transformations listed in (15). Thus, we note that all the dynamical variables 
(i.e. $x, \psi, \bar \psi$) acquire positive signature under the operations of  two successive discrete transformations (15).
We shall exploit this observation later on.

The nilpotent and 
continuous symmetry transformations $s_1$ and $s_2$ are connected by the following interesting relationship:
\begin{eqnarray}
s_2\; \Phi = \pm \;* \; s_1\; * \;\Phi , \qquad \quad \Phi = x, \;\psi, \;\bar  \psi,
\end{eqnarray} 
where the $(+)-$ signs on the r.h.s. of (17) are dictated by the signatures that are present in 
the relationship (15). One can also check that  
\begin{eqnarray}
s_1 \;\Phi = \mp \;* \;s_2 \;* \;\Phi , \qquad \quad \Phi = x, \;\psi, \;\bar  \psi,
\end{eqnarray} 
where we re-emphasize that the $(*)$, in 
equations (17) and (18), corresponds to the discrete transformations of equation (15).
The difference of signatures in (17) and (18) do crop up in theories with duality symmetry \cite{dght}.

We note that the relations in (17) and (18) are reminiscent of the relationship between the co-exterior derivative 
($\delta  = \pm *d*$) and the exterior derivative ($d = dx^\mu \partial_\mu$)
of differential geometry which are also nilpotent of order two 
(i.e. $d^2 = 0, \delta^2 = 0$). Thus, the Hodge duality $(*)$ operation of differential geometry is 
reflected in the existence of a set of discrete symmetry transformations (15)
for SHO. We also note that the duality operation $(*)$ on the fermionic conserved charges,
under (15), yields
\begin{eqnarray}
& * \; Q = \pm \;\bar Q, \qquad * \; \bar Q = \mp \;Q,& \nonumber\\
& *\; (*\; Q) = - \;Q, \qquad *\; (*\; \bar Q) = -\; \bar Q&.
\end{eqnarray} 
Thus, the conserved charges $Q$ and $\bar Q$ 
transform as: $ Q \to \pm \bar Q, \bar Q \to \mp Q$
which is like the duality transformations in the electrodynamics where we have : 
$\vec E \to \pm\vec B, \vec B \to \mp \vec E$ for the electric and magnetic fields present in the source-free Maxwell's equations.
Further, it should be noted that two successive operations of discrete transformations (15) on $Q$ and $\bar Q$
lead to $(-)$ sign on the r.h.s. This means that our present result is totally opposite to the 
effect of such operations on the dynamical
variables $x, \psi, \bar \psi$ [cf. (16)]. Finally, we find that (with $ * H = H, \; * W = - W, \; * N_B = - N_B,
* N_F = - N_F$), the total algebra amongst the operators $Q, \bar Q, W, N_B, N_F$ remains invariant under the
discrete symmetry transformations (15) which correspond to the $(*)$ operation.

\section{Algebraic structures}

One can very clearly verify that the continuous symmetry transformations ($s_1, s_2, s_W$) of {\bf section three} 
satisfy the following operator algebra:
\begin{eqnarray}
&&s_1^2 = 0, \quad s_2^2 = 0, \quad s_W = \{s_1, \; s_2\}  = (s_1 + s_2)^2, \nonumber\\
&&[s_1, \; s_W] = 0, \qquad [s_2, \; s_W] = 0, \qquad s_W^2 \ne 0.
\end{eqnarray}
The above algebraic structures are true on the on-shell condition where equations of motion 
$\ddot x + \omega^2 x = 0, \dot {\bar \psi} - i\omega \bar \psi = 0, \dot \psi + i \omega \psi = 0$ are satisfied.
The algebra (20) is reminiscent of the algebra obeyed by the de Rham cohomological operators 
$(d, \delta, \Delta)$ of 
differential geometry. The latter algebra, for the cohomological operators, can be succinctly 
expressed as \cite{smm,egh}
\begin{eqnarray}
&& d^2 = 0, \qquad \delta^2 = 0, \qquad \{d, \; \delta\}  = \;\Delta\; = (d + \delta)^2, \nonumber\\
&&[\Delta, \; d] = 0, \qquad [\Delta, \; \delta] = 0, \qquad \delta = \pm *\;d\;*,
\end{eqnarray}
where $\Delta$ is the Casimir operator and 
$(*)$ is the Hodge duality operation on a given spacetime manifold. The $(+)-$ signs in $\delta = \pm *d*$
are determined by the inner product of specific forms (see, e.g. \cite{smm,egh} for details). For our present model of SHO, we have 
already shown the analogue of relation ($\delta = \pm *d*$) in equations (17) and (18) for
the generic dynamical variable $\Phi$ of the theory.

One of the key properties of the exterior derivative $d$ is the fact that when it operates on a $n$-form ($f_n$)
it raises its degree by one (i.e. $d f_n \sim f_{n+1}$). Similarly, when $\delta$ acts on $f_n$, it lowers its degree by one
(i.e. $\delta f_n \sim f_{n-1}$). In contrast to $d$ and $\delta$, the Laplacian operator $\Delta$ does not change
the degree of a $n$-form ($f_n$), on which, it operates (i.e. $\Delta f_n \sim f_n$). These properties 
are very sacrosanct in the context of cohomological discussions.

The above properties are also captured within the purview of the symmetry considerations and 
conserved charges. To verify it, let us define the eigen state 
$|n_B \rangle$ with respect to the number operator $N_B$ (i.e. $N_B |n_B \rangle = n_B |n_B \rangle$).
Using the algebraic relations in (7), it can be explicitly checked  that  
\begin{eqnarray}
&& N_B \;Q\;|n_B \rangle = (n_B +1)\;Q \;|n_B \rangle, \nonumber\\
&&  N_B \;\bar Q\;|n_B \rangle = (n_B -1) \;\bar Q \;|n_B \rangle, \nonumber\\
&& N_B\; W\;|n_B \rangle = n_B\;W \;|n_B \rangle.
\end{eqnarray}
Thus, the eigen values of the states $Q|n_B\rangle, \bar Q|n_B\rangle$ and $W |n_B\rangle$, w.r.t. the operator $N_B$,
are $(n_B + 1), (n_B - 1)$ and $n_B$, respectively. This observation is similar to the consequences 
that ensue from the   operation of $d, \delta, \Delta$
on a given form of degree ($n_B$). Thus, the analogy between the
de Rham cohomological operators ($d, \delta, \Delta $) and the conserved charges 
($Q, \bar Q, W$) of our present theory is true.

The cohomological properties of ($d, \delta, \Delta $) can also be captured in terms of the fermionic number 
$n_F$ if we start with the eigen state $|n_F\rangle$ satisfying $N_F |n_F \rangle = n_F |n_F \rangle$. It is 
evident, from equations  (5) and (7), that 
\begin{eqnarray}
&&N_F \;\bar Q\;|n_F \rangle = (n_F +1)\;\bar Q \;|n_F \rangle, \nonumber\\
&& N_F \;Q\;|n_F \rangle = (n_F -1) \;Q \;|n_F \rangle, \nonumber\\
&&N_F\; W\;|n_F \rangle = n_F\;W \;|n_F \rangle.
\end{eqnarray}
We note, from the above,
 that the eigen values of the states $\bar Q|n_F\rangle,  Q|n_F\rangle$ and $W|n_F\rangle$,
w.r.t. the operator $N_F$,
are $(n_F + 1), (n_F - 1)$ and $n_F$, respectively. Thus, we conclude that there are 
two realizations of ($d, \delta, \Delta$)
in the language of symmetry generators ($Q, \bar Q, W$). If the degree of a form is identified with 
the bosonic number $n_B$,  we have the following mapping:
\begin{eqnarray}
(d, \;\delta, \;\Delta) \quad \Longleftrightarrow \quad (Q,\; \bar Q, \;W).
\end{eqnarray}
On the other hand, when the degree of a differential form is identified with $n_F$, then, 
the operations of ($d, \delta, \Delta$)
and symmetry generators lead to one-to-one correspondence as given below:
\begin{eqnarray}
(d,\; \delta, \;\Delta) \quad \Longleftrightarrow \quad (\bar Q, \; Q, \;W).
\end{eqnarray}
Thus, the algebraic structures of (7), in terms of the conserved operators, capture the algebraic structures of (21). As
a consequence, our present model (i.e. SHO) is a prototype example of the Hodge theory.

\section{Conclusions}
In our present investigation, we have shown that the Lagrangian (1) of 
SHO respects {\it three} continuous symmetry transformations (i.e. $s_1, s_2$ and $s_W$).
These continuous symmetries (and their corresponding generators) provide a physical 
realization of the de Rham cohomological operators
of  differential geometry at the algebraic level. In addition to the above continuous symmetries, 
the Lagrangian (1) also respects  discrete symmetries which correspond to the Hodge duality 
operation ($*$) of differential geometry. This symmetry enables us to 
establish the appropriate relations ($s_2 = \pm * s_1 *, s_1 = \mp * s_2 *$) between the 
transformations $s_1$ and $s_2$. These relations are exactly same as  
the relation ($\delta = \pm * d *$) between $\delta$ and $d$ of the differential geometry.

In addition to the above, the generator $Q$ (corresponding to the transformations $s_1$) 
increases  the bosonic number $n_B$ by one unit and decreases the fermionic number $n_F$  
by one unit whereas the generator $\bar Q$ (corresponding to the symmetry transformations $s_2$) 
does its opposite. The bosonic generator $W$ does not alter $n_B$ as well as $n_F$. 
It is interesting to point out that there is a  {\it one-to-one} correspondence between the 
generators ($Q, \bar Q, W$) and de Rham cohomological operators ($d, \delta, \Delta$) 
of differential geometry. The algebra obeyed by the symmetry transformations 
(and their corresponding generators) is reminiscent of the algebra of 
de Rham cohomological operators of differential geometry. 
Thus, the present theory provides a cute model for the Hodge theory.

The above Hodge algebra has also been obtained in the case of  2D free Abelian 1-form as well as 4D free 
Abelian 2-form gauge theories \cite{sr1,r,sr2}. In these theories, the nilpotent ($Q_{a(b)}^2 = 0$)
(anti-)BRST charges $Q_{a(b)}$ (corresponding to the (anti-)BRST symmetry transformations $s_{a(b)}$) 
increase and decrease the ghost number by one unit, respectively. The nilpotent 
($Q_{a(d)}^2 = 0$) (anti-)co-BRST charges (corresponding to the (anti-)co-BRST 
symmetry transformations $s_{a(d)}$)  have an opposite effect on the ghost number. The bosonic symmetry, 
obtained from the anticommutator of the above nilpotent symmetries, does not 
increase or decrease the ghost number. Therefore, there is  {\it two-to-one} mapping 
between the continuous symmetry transformations (and their corresponding generators) and 
the de Rham cohomological operators of differential geometry (see, e.g. \cite{r,sr1,sr2} for details).

It is clear from the above discussions that the symmetry transformations (and their corresponding generators) of SHO 
have similar kind of algebra as obeyed by the symmetry transformations (and their corresponding generators) of 
the 2D free Abelian 1-form as well as 4D free Abelian 2-form gauge theories. However, there is a glaring difference
as far as physical realizations of the 
cohomological operators are concerned. Whereas there is two-to-one mapping between
the conserved charges and the cohomological operators within the framework of BRST formalism, there is one-to-one
mapping between the conserved charges and cohomological operators in the case of SHO. 
Physically, we feel, the model of SHO is more appealing than the BRST analysis of  
gauge theories because the number operators ($n_F$ and $n_B$)
are more physical than the ghost number in the BRST approach. 
We plan to extend our present work to other superpotentials that have been discussed in \cite{cas}. In 
particular, we strongly feel that the theories with shape-invariant superpotentials would also be 
models for the Hodge theory.

In the existing literature on supersymmetric algebra and cohomological discussions \cite{jcb,rz,lf,fb}, 
a whole lot of developments have been made on the (hyper)K\"{a}hler manifolds. One dimensional 
supersymmetric structures have been found for the twisted and untwisted versions of the super Yang-Mills theory 
\cite{lf}. However, our present model of SHO is a simple case where the properties of the (super)K\"{a}hler manifolds  
are {\it not} invoked at all. The model of SHO is one of the simplest supersymmetric system that provides the physical 
realizations of cohomological operators in terms of suitable symmetries.

The analogue of the Laplacian operator ($\Delta $)
of differential geometry is the continuous symmetry transformation (12)
for SHO. However, as far as the basic tenet of supersymmetric quantum mechanics is concerned, the anticommutator
$\{ s_1, s_2 \}$ should lead to the time translation (i.e. $\{s_1, s_2\} \Phi \sim \dot \Phi$). It can be verified
that the transformations (12) satisfy this requirement, too, when we use the appropriate
equations of motion. To be precise,
it turns out that $\{s_1, s_2 \} \Phi = (i/\omega) \dot \Phi$ for the generic dynamical variable 
$\Phi = x, \psi, \bar \psi$ if we use the equations of motion $\dot \psi + i \omega \psi = 0$ and
$\dot {\bar \psi} - i \omega \bar \psi = 0$ [derived from the Lagrangian (1)]. Thus, our present model of SHO provides a 
complete and cute  model for the Hodge theory.

\vskip .04 cm
{\sf\large Acknowledgement:}
One of us (RK) would like to thank UGC, Government of India, New Delhi, for financial support under the JRF scheme.


\begin{thebibliography}{99}
\bibitem{w}    See, e.g., E. Witten, Nucl. Phys. B {\bf 185}, 531 (1981).
\bibitem{cas}  See, e.g., for an excellent review, F. Cooper, A. Khare, U. Sukhatme, Phys. Rep. {\bf 251}, 264 (1995).
\bibitem{smm}  See, e.g., S. Mukhi, N. Mukanda, {\it Introduction to Topology, Differential 
               Geometry and Group Theory for Physicists} (Wiley Eastern
               Private Limited, New Delhi, 1990).  
\bibitem{egh}  T. Eguchi, P. B. Gilkey, A. Hanson, \\Phys. Rep. {\bf 66}, 213 (1980).
\bibitem{r}    R. P. Malik, Mod. Phys. Lett. A {\bf 15}, 2079 (2000). 
\bibitem{sr1}  S. Gupta, R. P. Malik, Eur. Phys. J. C {\bf 58}, 517 (2008).
\bibitem{sr2}  S. Gupta, R. P. Malik, Eur. Phys. J C {\bf 68}, 325 (2010). 
\bibitem{rpm1} R. P. Malik, J. Phys. A: Math. Gen. {\bf 41}, 4167 (2001). 
\bibitem{rpm2} R. P. Malik, J. Phys. A: Math. Gen. {\bf 36}, 5095 (2003).
\bibitem{d}    P. A. M. Dirac, {\it Lectures on Quantum Mechanics} [Belfer Graduate
               School of Science] (Yeshiva University Press, New York, 1964). 
\bibitem{das}  See, e.g., for an excellent exposition, A. Das, {\it Field Theory: A Path Integral Approach}
               (World Scientific, Singapore, 1993). 
\bibitem{dght} S. Deser, A. Gomberoff, M. Henneaux, C. Teitelboim,\\
               Phys. Lett. B {\bf 400}, 80 (1997).
\bibitem{jcb}  J. M. Figueroa-O'Farrill, C. Koehl, B. Spence,    Nucl. Phys. B {\bf 503}, 614 (1997).              
\bibitem{rz}   R. Zucchini,  Class. Quant .Grav. {\bf 24}, 2073 (2007).
\bibitem{lf}   L. Baulieu, F. Toppan, Lett. Math. Phys. {\bf 98}, 299 (2011).
\bibitem{fb}   F. Brandt,  J. Math. Phys. {\bf 51}, 122302 (2010).
          
\end{thebibliography}
\end{document}